\documentclass[sigconf]{acmart}

\copyrightyear{2017}
\acmYear{2017}
\setcopyright{acmcopyright}
\acmConference{CIKM'17}{}{November 6--10, 2017, Singapore.}
\acmPrice{15.00}
\acmDOI{https://doi.org/10.1145/3132847.3132943}
\acmISBN{ISBN 978-1-4503-4918-5/17/11} 

\usepackage{booktabs} % For formal tables
\usepackage{multirow}
\usepackage{sistyle}
\SIthousandsep{,}
\usepackage{bbm}
\usepackage[british]{babel}
\usepackage{pgfplots}
\usetikzlibrary{patterns}
\usepackage{filecontents}

\definecolor{mercury}{RGB}{240,240,240}
\definecolor{gallery}{RGB}{250,250,250}
\definecolor{free_speech_aquamarine}{RGB}{0, 156, 114}
\definecolor{shakespeare}{RGB}{35, 184, 223}
\definecolor{flamingo}{RGB}{237, 88, 85}
\definecolor{c1}{RGB}{255,97,56}
\definecolor{c2}{RGB}{255,255,157}
\definecolor{c3}{RGB}{190,235,159}
\definecolor{c4}{RGB}{121,189,143}
\definecolor{c5}{RGB}{0,163,136}
\begin{document}

\title{Learning Visual Features from Snapshots for Web Search}

\author{Yixing Fan$^{\dagger, \ddag}$, Jiafeng Guo$^{\ddag}$, Yanyan Lan$^{\ddag}$, Jun Xu$^{\ddag}$, Liang Pang$^{\dagger, \ddag}$ and Xueqi Cheng$^{\ddag}$}
\affiliation{
  \institution{${\dagger}$University of Chinese Academy of Sciences\\$^{\ddag}$CAS Key Lab of Network Data Science and Technology, Institute of Computing Technology} 
}
\email{{fanyixing, pangliang}@software.ict.ac.cn, {guojiafeng, lanyanyan, junxu, xqc}@ict.ac.cn}

\renewcommand{\shortauthors}{F. Yixing et al.}

\begin{abstract}
When applying learning to rank algorithms to Web search, a large number of features are usually designed to capture the relevance signals. Most of these features are computed based on the extracted textual elements, link analysis, and user logs. However, Web pages are not solely linked texts, but have structured layout organizing a large variety of elements in different styles. Such layout itself can convey useful visual information, indicating the relevance of a Web page. For example, the query-independent layout (i.e., raw page layout) can help identify the page quality, while the query-dependent layout (i.e., page rendered with matched query words) can further tell rich structural information (e.g., size, position and proximity) of the matching signals. However, such visual information of layout has been seldom utilized in Web search in the past. In this work, we propose to learn rich visual features automatically from the layout of Web pages (i.e., Web page snapshots) for relevance ranking. Both query-independent and query-dependent snapshots are considered as the new inputs. We then propose a novel visual perception model inspired by human's visual search behaviors on page viewing to extract the visual features. This model can be learned end-to-end together with traditional human-crafted features. We also show that such visual features can be efficiently acquired in the online setting with an extended inverted indexing scheme. Experiments on benchmark collections demonstrate that learning visual features from Web page snapshots can significantly improve the performance of relevance ranking in ad-hoc Web retrieval tasks.

\end{abstract}



% We no longer use \terms command


\maketitle

\begin{figure*}[!tbp]
\centering
\includegraphics[scale=0.65]{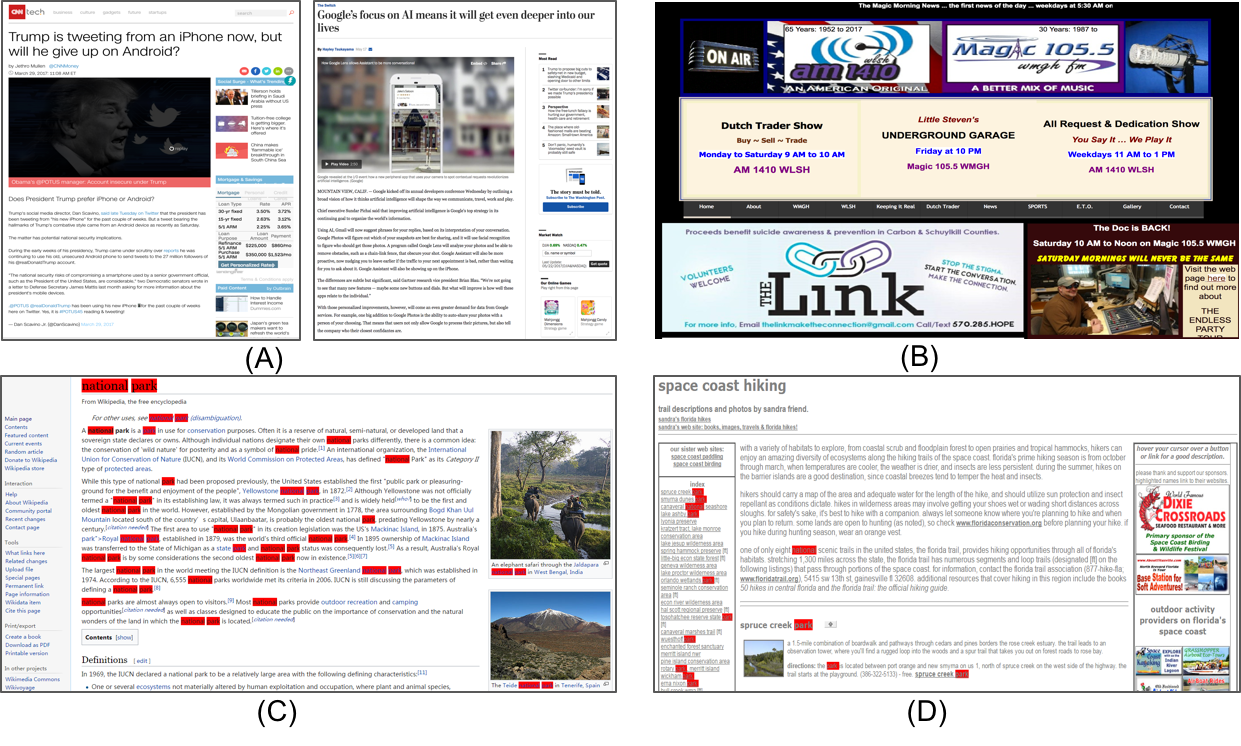}
\caption{Snapshots of different kinds of Web page: (A) High quality Web pages with formal layout; (B) Low quality Web pages with images and advertisement; (C) A relevant page of the query ``national park'' with keyword matching highlighted; and (D) A non-relevant page of the query ``national park'' with keyword matching highlighted.}
\label{fig:motivation}
\end{figure*}

\section{Introduction}
Modern search engines have widely adopted learning to rank (LTR) methods for Web page ranking. A fundamental step of LTR methods is to design a large number of features which are capable of characterizing the relevance between a document and a query. As revealed in literature, most of these features are computed based on the extracted textual elements (e.g., title, main content, and anchor texts), link analysis (e.g., PageRank and HITS) and user logs (e.g., clickthrough ratio). For instance, in the well-known LETOR $4.0$ collection~\cite{qin2010letor}, there are $46$ human-crafted features in total, among which $42$ features are constructed based on the textual elements (e.g., term frequencies, BM25 and language model scores based on title, body, anchor texts, and URL), and 4 features are based on link analysis (e.g., PageRank, inlink number, outlink number, and number of child page).

However, Web pages are not solely linked texts, but have structured layout organizing a large variety of elements in different styles. Such layout itself can convey useful visual information, indicating the relevance of a Web page.
In the first place, the query-independent layout, i.e., the raw Web page layout, can help identify the page quality. For example, a high-quality Web page of news, blog or review is often well structured with head bars, side bars, and main body containing rich textual content, as shown in Figure~\ref{fig:motivation} (A). Users can perceive the authority of a page from its formal layout. On the contrary, a low-quality Web page may contain many floating images and advertisement with useless information, as shown in Figure~\ref{fig:motivation} (B).  Secondly, the query-dependent layout, i.e., the Web page rendered with matched query words, can further tell rich structural information of the matching signals. For example, on a relevant Wikipedia page of the query ``national park'' as shown in Figure~\ref{fig:motivation} (C), we can observe that there are many matching signals distributed in the main content, large matching signals in the title, and high spacial proximity between these matching signals. On the contrary, an irrelevant Web page of ``national park'' could also contain a number of query words but in which the matching signals may distribute in side areas (e.g., advertisement) as shown in Figure~\ref{fig:motivation} (D), or even be invisible in some spamming pages due to keyword stuffing. In summary, search users may perceive many useful visual information from the Web page layout for relevance judgment, while such information have not been effectively modeled in Web search in literature.

There have been a few studies attempting to take into account the layout information in Web search but from a non-visual way. For example,
Zhou et al.~\cite{zhou2005document} observed that pages like tables and lists are unlikely to be relevant for ad hoc queries, and assumed such pages have unusual word distributions. Based on this assumption, they constructed two content features to estimate the Web page quality.
Bendersky et al.~\cite{bendersky2011quality} designed page quality features relate to the readability, layout, ease-of-navigation and so on from HTML tags and textual content. These features are used to promote high-quality pages and penalize low-quality pages in Web search.
Obviously, all the above methods utilized the layout information indirectly (i.e., defined from textual content or HTML tags) with manually designed features. This may largely restrict the exploitation of the visual information in Web page for relevance modeling.

In this paper, we propose to learn rich visual features automatically from the layout of Web pages for relevance ranking. Specifically, we take the snapshot of a Web page (i.e., a rendered image of the Web page) as a new type of input in the learning to rank framework. Both query-independent and query-dependent snapshots have been introduced in our work. We then propose a \textit{visual perception} (ViP) model inspired by human's visual searching behaviors on page viewing (i.e., F-biased viewing pattern~\cite{Faraday2000}) to extract visual features from the snapshots for relevance ranking. Specifically, the ViP model is a neural model which contains four stacked layers, namely snapshot segmentation layer, local perception layer, sequential aggregation layer, and relevance decision layer. The proposed ViP model can be learned in an end-to-end way together with traditional human-crafted features. Besides, for practical implementation of the ViP model, we also introduce an efficient indexing scheme for Web page snapshots.

We evaluate the effectiveness of the proposed model based on two representative ad-hoc retrieval benchmark datasets from the LETOR collection~\cite{qin2010letor}. For comparison, we take into account some well-known traditional retrieval models as well as several state-of-the-art learning to rank models. The empirical results show that our model outperform all the baselines in terms of all the evaluation metrics.
We also provide detailed analysis on the proposed model, and conduct case studies to provide better understanding on the learned visual signals.

The main contributions of this paper include:
\begin{description}
\item{1.} We argue that visual information from Web page layout is valuable for relevance modeling in Web search, and introduce Web snapshots as an additional input for learning visual features.
\item{2.} We propose a novel visual perception model over the Web snapshots, which can automatically learn visual features for relevance modeling in an end-to-end way.
\item{3.} We conduct rigorous comparisons over existing representative retrieval models, and demonstrate that learning visual features from Web snapshots can significantly improve the performance of relevance ranking in ad-hoc Web retrieval tasks.
\end{description}

\section{Related Work}
In this section, we briefly review three research areas related to our work, including visual search, Web page quality based on layout and Web search using layout information.

\subsection{Visual Search}
Visual search, which studies how visual elements affect users' information seeking experience and how users view Web pages, has been extensively studied in the past decades \cite{scott1993visual, wolfe1994guided, kitajima2000comprehension}.

Although Web pages mainly rely on textual content to convey information, page layout has been recognized as of great importance in information seeking experience\footnote{https://www.nngroup.com/articles/let-users-control-font-size/}. It is imperative that Web pages are constructed to enable a high level of usability for all users~\cite{shneiderman2000universal}, but poorly designed layouts can quickly lead to fatigue, with a resultant lowering of speed and accuracy on task performance~\cite{streveler1984quantitative}.
In~\cite{larson1998web}, Larson et al. studied the principles for the design of multiple hyperlinks on a Web page for information retrieval tasks. They showed a medium structure in terms of depth and breadth outperformed the broadest but shallow structure overall.
In~\cite{Ling2002The}, Ling et al. studied the effect of the combination of text and background color on visual search performance and subjective preference. They found that higher contrasts between text and background color led to faster searching and were rated more favourable.
Ling~\cite{ling2006influence} explored the influence of the font type and line length on two tasks, i.e., visual search and information retrieval. They found the effect of line length was significant, while font type has little impact on task performance.
Pearson et al.~\cite{pearson2003effect} studied the effect of spatial layout and link color in Web pages on performance of visual search and interactive search tasks. They found that both link color and presentation position of menus have significant effect on user's information seeking experience.

Besides these above analysis, there have also been a line of studies on how users view Web pages. It has been widely accepted that users' Web-viewing behavior is significantly different from that on natural images. Specifically, several distinct patterns have been revealed in the past work,  such as F-biased viewing pattern which scans a page in ``F'' shape~\cite{Faraday2000, scott1993visual}, and banner blindness pattern which avoids banner-like advertisement~\cite{grier2007users}.

These previous studies have shown that the layout of a Web page has significant impact on users' information seeking experience and also inspired us on model designing for learning visual features from Web page layouts.

\subsection{Web Page Quality based on Layout}
In the field of Human-Computer Interaction (CHI), it has been widely studied how different layouts of Web pages affect users' quality decisions on Web pages. 

In~\cite{fogg2001makes}, Fogg et al. conducted an online study that investigated how different elements of Web sites affect people's perception of credibility. They found seven types of elements, where five types increase credibility perceptions and two hurt credibility.
Fogg et al.~\cite{fogg2003users} gathered \num{2684} people's comments about the evaluation on credibility of two live Web sites. They found the ``design look'' of the Web page the most prominent issue when people evaluated Web page credibility.

Besides the above questionnaire methods, there have been a line of studies on the effect of the layout based on automatic analysis.
One of the first automatic assessment systems originated in Web engineering. Syntax checkers were employed over HTML codes to analyze the quality of Web pages~\cite{bowers1996weblint}.
Chakrabarti et al.~\cite{chakrabarti2002structure} introduced six features of Web pages, such as the dominant color, the presence of advertisement, logos, animations, frames, and the frequency of links and graphics, to analyze the quality of Web pages. They found that pages which follow popular Web design guidelines might attract more viewers than other pages.
Mandl~\cite{mandl2006implementation} extracted about 100 features of Web pages by a page profiler as indicators of the quality. These features are mainly derived from the HTML code and try to capture design aspects, for example, number of list, number of colors, number of DOM elements and so on.
Song et al.~\cite{song2004learning} utilized a vision-based page segmentation algorithm to partition a Web page into semantic blocks based on the visual layout information. Spatial features (e.g., position and size) and content features (e.g., the number of images and links) were extracted to estimate the importance of each blocks.

All these studies demonstrated that the Web page layout has strong impact on users' perception of the Web page quality.

\begin{figure*}[!tbp]
\centering
\includegraphics[scale=0.58]{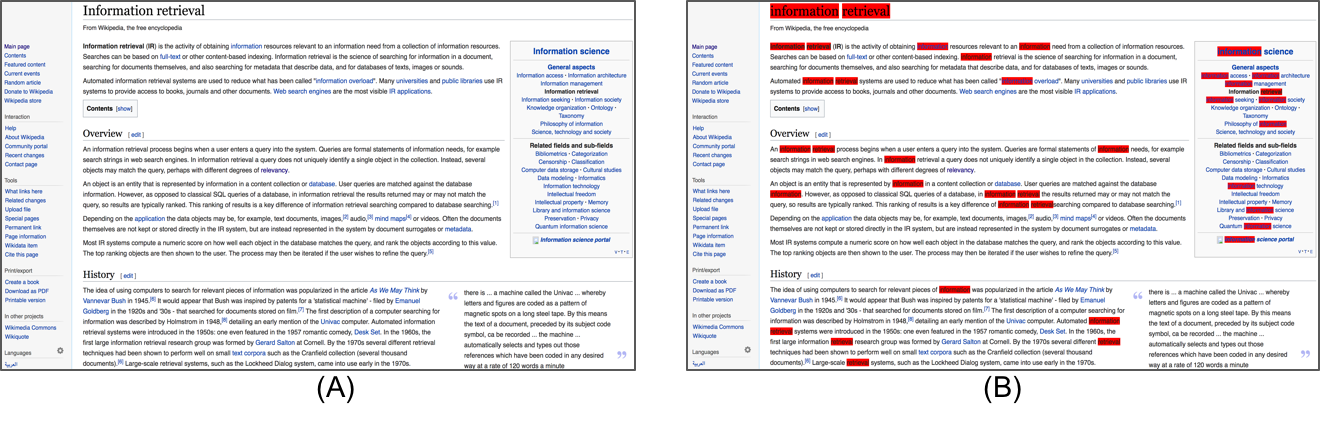}
\caption{(A) Query-independent snapshot. (B) Query-dependent snapshot.}
\label{fig:render}
\end{figure*}

\subsection{Web Search using Layout Information}
There have been a few studies attempting to take into account the layout information in Web search. For example,
Zhou et al.~\cite{zhou2005document} observed that Web pages like tables and lists are unlikely to be relevant for ad hoc queries, and assumed such pages have unusual word distributions. Based on this assumption, they constructed two content features to estimate the Web page quality. They incorporated the page quality into language model and demonstrated that it can significantly outperform the query likelihood model.
Bendersky et al.~\cite{bendersky2011quality} presented a quality-biased ranking method that promotes Web pages containing high-quality content, and penalizes low-quality Web pages. In their model, the quality of the page content is determined by the readability, layout, ease-of-navigation and so on. Their results showed that by taking into account the quality of the Web page, consistent retrieval performance improvement could be obtained as compared with the methods relying on text-based and link-based features.
Pirlo et al.~\cite{pirlo2013layout} presented a layout-based retrieval system to search commercial forms.
They constructed layout-based features from the extracted grid-based structural components. They demonstrated the effectiveness of the layout information in retrieval of commercial forms.

Although the Web page layout has been considered in retrieval in previous work, the existing models usually utilized the layout information indirectly (defined from textual content or HTML tags) with manually designed features. This may largely restrict the exploitation of the visual information in Web page layout for relevance modeling.

\section{Our Approach}
In this section, we describe our model on learning visual features from Web page layout for relevance ranking in detail. Specifically, we take the snapshot of a Web page as the input, and propose a visual perception (ViP) model to extract visual features from the snapshots. This model is learned end-to-end together with traditional human-crafted features. In the following, we first introduce the snapshot construction process. We then talk about the ViP model and model training in detail. Finally, we discuss a new indexing scheme for the implementation of our ViP model.

\subsection{Snapshot Construction}
Typically, a Web page is a document written in HTML or comparable markup language, organizing various  Web resource elements in a structured way. Web browsers coordinate the various elements for the written page to present the Web page to users. In order to leverage the layout information of Web pages, we propose to render the source Web page into a snapshot as is shown in Web browsers perceived by search users. This render process could be efficiently conducted using an simple render tool.

In our work, we consider two types of snapshots for a Web page, namely query-independent snapshot and query-dependent snapshot. The query-independent snapshot captures the raw Web page layout information, which can be directly generated using the render tool over the raw Web page source code, as illustrated in Figure~\ref{fig:render} (A) . The query-dependent snapshot, on the other hand, aims to capture the Web page layout information as well as matching signals given a specific query. This is to simulate how users perceive a Web page given the information need. We achieve this by highlight the matched query words on a Web page using some background color, as shown in Figure~\ref{fig:render} (B) . In this work, all the query words are rendered with the same background color for simplicity. In fact, one may use different colors for different words to convey richer information (e.g., query word importance) and we will leave this as our future work.

\subsection{The Visual Perception Model}
Given the snapshot of a Web page, here we aim to design a model that can learn visual features automatically for relevance ranking. As the snapshot is an image, a simple idea is to directly employ an existing neural model, e.g., the convolutional neural network (CNN), for this purpose. However, users' viewing patterns on Web pages may not be the same as that on the general image~\cite{shen2014webpage}. A model that can better fit users' viewing patterns on Web pages may lead to better feature learning performance on the snapshots.

\begin{figure*}[!tbp]
\centering
\includegraphics[scale=0.55]{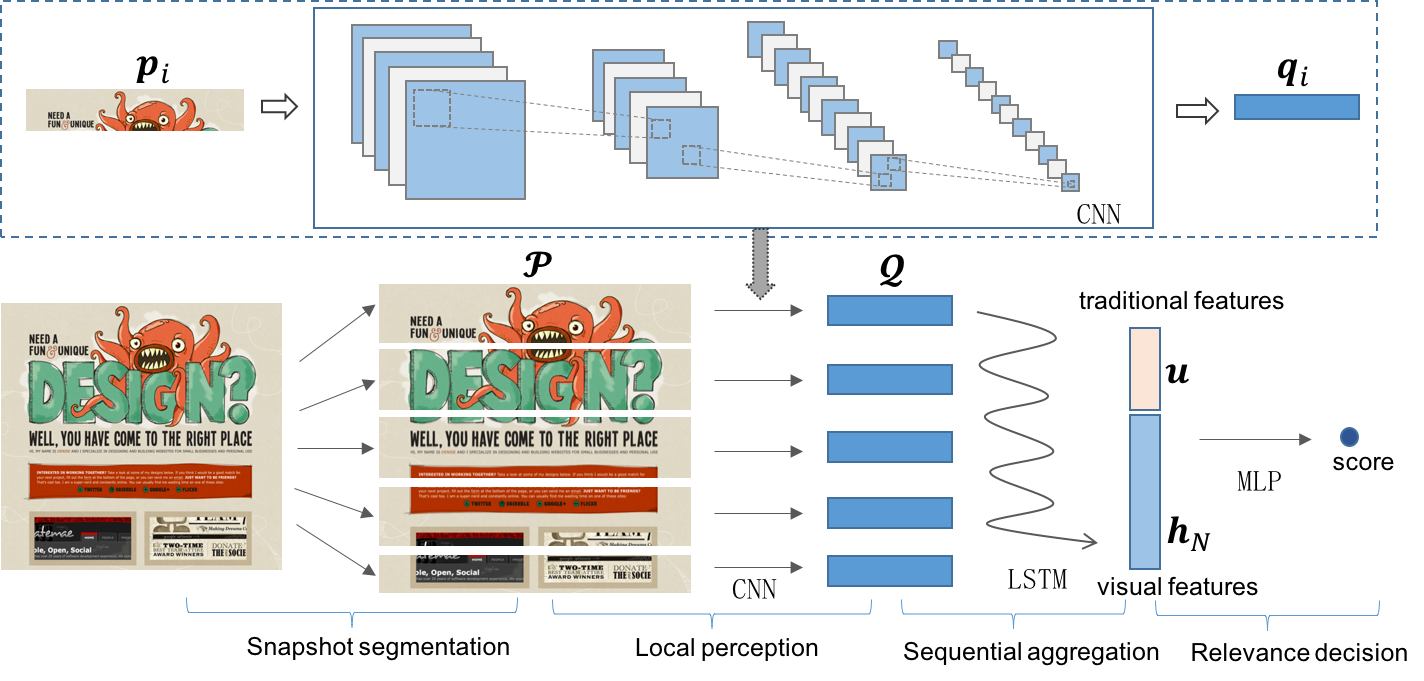}
\caption{The Architecture of the ViP Model}
\label{fig:vip-model}
\end{figure*}

In fact, there have been extensive studies on how users view Web pages in the field of visual search~\cite{Faraday2000, kitajima2000comprehension, scott1993visual}. It has been widely accepted that users are accustomed to reading row by row from top to bottom, which forms the well-known F-biased viewing pattern~\cite{Faraday2000}. Inspired by these observations, we propose a deep neural model that can simulate the F-biased viewing pattern of search users on Web pages to extract visual features from snapshots. We refer to our model as a visual perception (ViP) model. The architecture of the ViP model is depicted in Figure~\ref{fig:vip-model}, which contains four stacked layers, namely snapshot segmentation layer, local perception layer, sequential aggregation layer, and relevance decision layer. In the following, we will introduce these layers in detail.

\subsubsection{Snapshot Segmentation Layer}
The snapshot segmentation layer focuses on producing a set of region proposals. This is a typical step for image processing and different region segmentation methods have been proposed in different tasks, such as selective search~\cite{uijlings2013selective}, objective detection~\cite{girshick2014rich, cirecsan2013mitosis}, multi-scale combinatorial grouping~\cite{arbelaez2014multiscale} and so on. In this work, we propose to generate a set of horizontal region proposals to simulate the row by row scanning behaviors in F-biased viewing pattern. Specifically, we segment a snapshot into several horizontal regions with equal height as shown in Figure~\ref{fig:vip-model}. Different heights actually capture different granularity in row scanning, and we have studied the height effect in Section 4.4. Formally, given an input snapshot image $I$, a set of region proposals $\mathcal{P} = \{p_1, p_2, ..., p_N\}$ are generated, where $N$ denotes the number of region proposals.

\subsubsection{Local Perception Layer}
Based on the above region proposals, we employ a convolutional neural network, which is good at producing abstract image features, to generate row features from each local region. Specifically, for each region proposal $p_i$, the $k-$th kernel $\mathbf{W}^{(1,k)}$ scans over the proposal $\mathbf{Z}^{(0)} = p_i$ to generate a feature map $\mathbf{Z}^{(1,k)}$:
	\begin{equation}
		\mathbf{Z}_{i,j}^{(1,k)} = \sigma(\sum_{s=0}^{r_k-1}\sum_{t=0}^{r_k-1}\mathbf{W}_{s,t}^{(1,k)}\cdot \mathbf{Z}_{i+s, j+t}^{(0)} + \mathbf{b}^{(1,k)})\,,
	\end{equation}
	where $r_k$ denotes the size of the $k-$th kernel. In this paper, we use square kernel, and adopt ReLU~\cite{Dahl2013Improving} as the activation function $\sigma$. $\mathbf{W}$ and $\mathbf{b}$ are parameters to be learned. We take a max-pooling after each convolution:
	\begin{equation}
		\mathbf{W}_{i,j}^{(2,k)} = \max_{0 \leq s \le d_k}\max_{0 \leq t \le d_k} \mathbf{Z}_{i \cdot d_k + s, j \cdot d_k + t}^{(1,k)}\,,
	\end{equation}
	where $d_k$ denotes the width of the pooling kernel.
	
	After the first convolution and max pooling layer, we continue to obtain higher abstract features $\mathbf{Z}^{(l)}$, $l \geq 2$ by further convolution and max pooling, with general formulations:
	\begin{equation}
		\mathbf{Z}_{i,j}^{(l+1, k^{'})} = \sigma\big(\sum_{k=0}^{c_l-1}\sum_{s=0}^{r_k-1}\sum_{t=0}^{r_k-1} \mathbf{W}_{s,t}^{(k+1, k^{'})} \cdot \mathbf{Z}_{i+s, j+t}^{(l,k)} + \mathbf{b}^{(l+1, k)} \big), l = 2, 4, 6 ...
	\end{equation}
	\begin{equation}
		\mathbf{Z}_{i,j}^{(l+2,k^{'})} = \max_{0 \leq s \le d_k} \max_{0 \leq t \le d_k} \mathbf{Z}_{i\cdot d_k + s, j \cdot d_k + t}^{(l+1,k^{'})}\,,
	\end{equation}
	where $c_l$ denotes the number of feature maps in the $l-$th layer.
	
	Finally, the output of the last max pooling layer $\mathbf{Z}^{(l)}$ is flattened as the row feature $\mathbf{q}_i$ of the local region proposal $p_i$.

\subsubsection{Sequential Aggregation Layer}
Based on the row features $\mathbf{Q} = \{ \mathbf{q}_1, \mathbf{q}_2, \mathbf{q}_3, ..., \mathbf{q}_N\}$ generated in the local perception layer, we attempt to generate the overall visual features by aggregating these row features.
We adopt the recurrent neural network which naturally fits the sequentially scanning behaviors (i.e., from top to bottom) in F-biased viewing pattern.
Here, we use long-short term memory network (LSTM)~\cite{hochreiter1997long}, a powerful model for variable-length sequential data, to accumulate the features $\mathbf{q}_i$ from each local region. Specifically, as shown in Figure \ref{fig:vip-model}, we feed the row features into LSTM sequentially to generate the accumulated features at different positions as follows.
   \begin{align}
	  %\begin{split}
		&  \mathbf{i}_t = \sigma(\mathbf{W}_{i}\mathbf{q}_t + \mathbf{U}_{i}\mathbf{h}_{t-1} + \mathbf{b}_i)\,, \\
		&  \mathbf{f}_t = \sigma(\mathbf{W}_{f}\mathbf{q}_t + \mathbf{U}_{f}\mathbf{h}_{t-1} + \mathbf{b}_f)\,, \\
		&  \mathbf{c}_t = \mathbf{f}_t\mathbf{c}_{t-1} + \mathbf{i}_t\tanh(\mathbf{W}_{c}\mathbf{q}_t + \mathbf{U}_{c}\mathbf{h}_{t-1}+\mathbf{b}_c)\,, \\
		&  \mathbf{o}_t = \sigma(\mathbf{W}_{o}\mathbf{q}_t + \mathbf{U}_{o}\mathbf{h}_{t-1} + \mathbf{b}_o)\,, \\
		&  \mathbf{h}_t = \mathbf{o}_t\tanh(\mathbf{c}_t)\,,
	  %\end{split}
	\end{align}
where $\mathbf{q}_t$ denotes the $t$-th proposal features, $\mathbf{i}_t$,$\mathbf{f}_t$,$\mathbf{o}_t$ denote the input, forget, and output gates respectively, $\mathbf{c}_t$ denotes the information stored in memory cell and $\mathbf{h}_t$ denotes the $t$-th accumulated evidence, $\mathbf{W}_i$,$\mathbf{W}_f$,$\mathbf{W}_o$, $\mathbf{W}_c$, $\mathbf{U}_i$,$\mathbf{U}_f$,$\mathbf{U}_o$,$\mathbf{U}_c$,$\mathbf{b}_i$,$\mathbf{b}_f$,$\mathbf{b}_o$ and $\mathbf{b}_c$ are parameters to be learned, $\sigma$ denotes the sigmoid function.

We take the last output of the LSTM model as the visual features $h_N$ of the snapshot.

\subsubsection{Relevance Decision Layer}
To generate the final relevance score of a Web page, we leverage both the extracted visual features from the snapshot and traditional human-crafted features. Specifically, we concatenate the visual feature vector $h_N$ with the traditional feature vector $\mathbf{u}$ to form the final relevance features $\mathbf{v}$. We then feed the relevance feature $\mathbf{v}$ into a 2-layer feedforward neural network to get the final relevance score $\mathbf{s}$.
\begin{equation}
	\mathbf{s} = \mathbf{W}^1 \cdot \sigma( \mathbf{W}^0 \cdot \mathbf{v} + \mathbf{b}^0) + \mathbf{b}^1\,,
\end{equation}
where $\mathbf{W}^0$, $\mathbf{W}^1$, $\mathbf{b}^0$, and $\mathbf{b}^1$ are parameters to be learned, $\sigma$ denotes the ReLU function. In this way, the ViP model can be learned end-to-end to automatically extract visual features from Web snapshots as well as to produce a relevance model.

\subsection{Model Training}
 Since the ad-hoc Web retrieval task is fundamentally a ranking problem, we utilize the pairwise ranking loss such as hinge loss to train our model. Specifically, given a triple $(q, d^+, d^-)$, where $d^+$ is ranked higher than $d^-$ with respect to query $q$, the hinge loss function is defined as:
$$\mathcal{J}(q, d^+, d^-; \theta) = \max(0, 1-s(q, d^+) + s(q,d^-))\,,$$
where $s(q,d)$ denotes the relevance score for $(q,d)$, and $\theta$ includes the parameters in the local perception layer, sequential aggregation layer, and relevance decision layer.
It is worth noting that the overall model is a combination of traditional human-crafted features and learned visual features, where the traditional features are static without learning in this model. In this way, the model can easily be biased to the visual features. Thus, we introduced $\ell_2$ regularization terms into the loss function,
	$$\mathcal{L}(q, d^+, d^-; \theta) = \mathcal{J}(q, d^+, d^-; \theta) + \lambda_1 \left \lVert \Phi_1 \right \rVert_2^2 +
	\lambda_2 \left \lVert \Phi_2 \right \rVert_2^2\,,
	$$
	where $\Phi_1$ denotes parameters in the local perception layer and sequential aggregation layer, $\Phi_2$ denotes parameters in the relevance decision layer, $\lambda_1$ and $\lambda_2$ denote the corresponding regularizer co-efficient, respectively. In this way, we can introduce stronger regularization on $W_1$ to alleviate the model bias.
	The optimization is relatively straightforward with standard backpropagation. We apply stochastic gradient decent method Adam \cite{kingma2014adam} with mini-batches (100 in size), which can be easily parallelized on single machine with multi-cores.

\begin{figure*}[!tbp]
\centering
\includegraphics[scale=0.5]{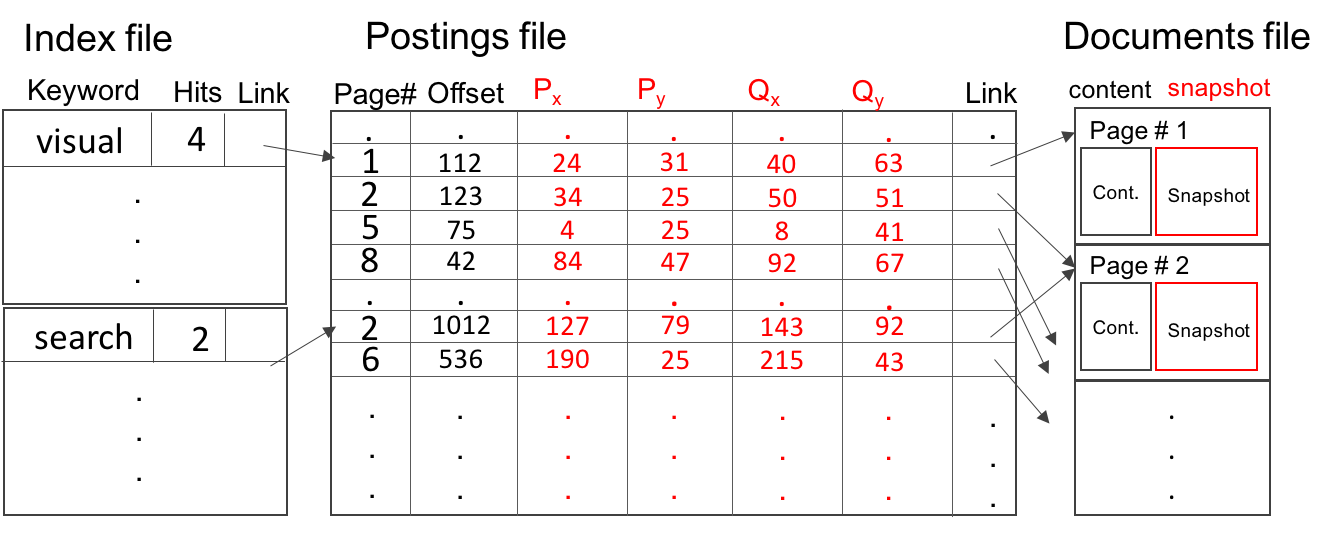}
\caption{The indexing scheme to incorporate snapshot information into the inverted files.}
\label{fig:index_rendered}
\end{figure*}

\subsection{The Indexing Scheme of Snapshots}
In order to implement the ViP model for practical Web search, we need an efficient indexing scheme for the Web page snapshots. For query-independent snapshots, it is simple to implement since we only need to associate each Web page snapshot to its page ID. For query-dependent snapshots, we propose an efficient indexing scheme that can be well incorporated into the widely adopted keyword based inverted indexing for implementation. Specifically, during the traditional inverted indexing construction process, for each keyword in a Web page, we generate a keyword-dependent snapshot and recognize all the highlighted positions of this keyword in the snapshot. Each highlighted position is actually a rectangle, so we use the position of the top-left and bottom-right of the rectangle to record it. Specifically, we record the relative offsets of these two positions, denoted by $(P_x,P_y)$ and $(Q_x,Q_y)$ in Figure~\ref{fig:index_rendered}, with respect to the top-left corner of the snapshot. We append these offsets to the posting files after the original position of the keyword in the page, denoted by $\mathit{Offset}$. In this way, the keyword-dependent snapshot can be discarded since all the highlighted positions have been recorded.

At the testing time, given a query and a candidate Web page, we can obtain all the highlighted positions in the snapshot of all the query keywords during the posting list merging process. We can also obtain the query-independent snapshot of the Web page by the page ID. In this way, we can easily generate the query-dependent snapshot by simply modifying the corresponding pixels at all the highlighted positions to some predefined color based on the query-independent snapshot.

\section{Experiment}
In this section, we conduct experiments to demonstrate the effectiveness of our proposed model on benchmark collections.

\begin{table}[bpt]
\centering
\caption{Statistics of the datasets used in this study.}
\begin{tabular}{c c c c c c}
\hline
 & \#queries & \#pages & \#q\_rel &  \#rel\_per\_q\\ \hline
 MQ2007 & \num{1692} & \num{65323} & \num{1455} & 10.3 \\
 MQ2008 & \num{784} & \num{14384} & \num{564} & 3.7 \\ \hline
\end{tabular}
\label{tab:data_sets}
\end{table}

\subsection{Experimental Settings}
We first introduce our experimental settings, including datasets, baseline methods/implementations, and evaluation methodology.

\subsubsection{Data Sets}
To evaluate the performance of our model, we conducted experiments using two LETOR benchmark datasets~\cite{qin2010letor}: Million Query Track 2007 (MQ2007) and Million Query Track 2008 (MQ2008).
Both datasets use the GOV2 collection which includes 25 million Web pages in \num{426} gigabytes. We choose these two datasets according to three criteria: 1) the dataset is public; 2) there are a large number of queries compared with other public datasets; 3) the source Web page is available.
The details of the two datasets are given in Table~\ref{tab:data_sets}. As we can see, there are \num{1692} queries on MQ2007 and \num{784} queries on MQ2008. However, the number of queries with at least one relevant page is \num{1455} and \num{564}, respectively. The average number of relevant page per query is about \num{10.3} and \num{3.7} on MQ2007 and MQ2008, respectively.
In LETOR, there are \num{46} human-crafted features for each query and document pair, as described in the Introduction section.

\subsubsection{Snapshot Pre-Processing}
For each Web page, we generated both query-independent and query-dependent snapshots using a render tool. During rendering, we found \num{180} pages in total failed to generate the snapshots due to the missing link objects (e.g., stylesheet) over the two collections. For these pages, we employed a fake snapshot by averaging all other snapshots. It is worthy noting the length of the snapshots may vary significantly over different Web pages. To make it simple and efficient, we only kept the first screen size of the snapshot and down-sampled it to fixed resolution (i.e., \num{64}$\times$\num{64}), which corresponds to the first impression of the page for search users. We have also studied the performance of snapshots at different resolutions in Section 4.5. Besides, a simple image normalization was conducted by removing the average pixel values per data point, and then re-scaling linearly the range to [-1, 1].

\begin{table*}[!ht]
\centering
\caption{Analysis of the ViP model over the MQ2007 and MQ2008 datasets. Significant improvement or degradation with respect to $ViP_{\mathit{Baseline}}$ is indicated (+/-) ($\text{p-value} \le 0.05 $).}
\begin{tabular}{c c c c c c c c l}
\multicolumn{9}{c}{MQ2007}\\
\hline\hline
 & Model Name & P@1 & P@5 & P@10 & NDCG@1 & NDCG@5 & NDCG@10 & MAP\\
\hline
 without snapshot & ViP$_\mathit{Baseline}$ & 0.478~~ & 0.416~~ & 0.386~~ & 0.410~~ & 0.415~~ & 0.445~~ & 0.467~~ \\
 \hline
 \multirow{2}{3cm}{\centering query independent snapshot} & ViP$_\mathit{CNN}$ & 0.482~~ & $0.428^+$ & 0.390~~ & $0.418^+$ & $0.427^+$ & 0.451~~ & 0.472~~ \\
  & ViP & $0.494^+$ & $0.435^+$ & $0.397~^+$ & $0.425^+$ & $0.436^+$ & $0.461^+$ & 0.476 \\
\hline
 \multirow{2}{3cm}{\centering query dependent snapshot} & ViP$_\mathit{CNN}$ & 0.486~~ & $0.430^+$ & 0.393~~ & $0.421^+$ & $0.430^+$ & 0.453~~ & 0.475~~ \\
  & ViP & $0.505^+$ & $0.439^+$ & $0.398^+$ & $0.434^+$ & $0.441^+$ & $0.464^+$ & $0.481^+$ \\
 \hline\hline
 \multicolumn{9}{c}{MQ2008}\\
\hline\hline
%\toprule
 & Model Name & P@1 & P@5 & P@10 & NDCG@1 & NDCG@5 & NDCG@10 & MAP\\
 \hline
 without snapshot & ViP$_\mathit{Baseline}$ & 0.437~~ & 0.340~~ & 0.248~~ & 0.365~~ & 0.472~~ & 0.228~~ & 0.473~~ \\
 \hline
 \multirow{2}{3cm}{\centering query independent snapshot}  & ViP$_\mathit{CNN}$ & $0.449^+$ & 0.343~~ & 0.249~~ & 0.372~~ & 0.473~~ & 0.229~~ & 0.477~~\\
 & ViP & $0.458^+$ & 0.346~~ & 0.250~~ & $0.382^+$ & 0.475~~ & 0.230~~ & 0.480~~ \\
 \hline
 \multirow{2}{3cm}{\centering query dependent snapshot}  & ViP$_\mathit{CNN}$ & $0.454^+$ & 0.346~~ & 0.249~~ & $0.375^+$ & 0.474~~ & 0.229~~ & 0.479~~\\
 & ViP & $0.466^+$ & $0.356^+$ & $0.252^+$ & $0.396^+$ & $0.494^+$ & $0.235^+$ & $0.494^+$ \\
 \hline\hline
 %\bottomrule
\end{tabular}
\label{tab:structural_results}
\end{table*}

\subsubsection{Baseline Methods}
We adopt two types of baselines for comparison, including traditional retrieve models and the state-of-the-art learning to rank models.
Traditional retrieval models include
\begin{description}
	\item \textbf{QL}: Query likelihood model is one of the best performing language models based on Dirichlet smoothing~\cite{zhai2001study}.
	\item \textbf{BM25}: The BM25 formula~\cite{robertson1994some} is another highly effective retrieval model that represents the classical probabilistic retrieval model.
\end{description}
Learning to rank models include
\begin{description}
	\item \textbf{RankSVM}: RankSVM~\cite{joachims2006training} is a representative pairwise learning to rank model based on SVM$^{\text{struct}}$.
	
	\item \textbf{RankBoost}:RankBoost~\cite{freund2003efficient} formalizes learning to rank as a problem of binary classification, and combines a set of weak rankers as final ranking function based on boosting approach.
	\item \textbf{AdaRank}: AdaRank~\cite{xu2007adarank} is another boosting approach which aims to directly optimize the performance measure. Here we utilize NDCG as the performance measure function.
	\item \textbf{LambdaMart}: LambdaMart~\cite{burges2010ranknet} is a state-of-the-art learning to rank algorithm that uses gradient boosting to produce an ensemble of retrieval models.
\end{description}
For RankSVM, we directly use the implementation in $\mathit{SVM}^{\mathit{rank}}$~\cite{joachims2006training}. RankBoost, AdaRank, and LambdaMart are implemented using RankLib\footnote{https://sourceforge.net/p/lemur/wiki/RankLib/}, which is a widely used learning to rank tool.

We refer to our proposed model as \textbf{ViP}. For network configurations (e.g., numbers of layers and hidden nodes), we tune the hyper-parameters on a validation set.
Specifically, in the local perception layer, we set the size of region proposal to $4 \times 64$. We have also studied the performance of different proposal sizes in Section 4.4. For each proposal, there are 2 convolution layer each followed by a max pooling layer. In the first convolution layer, there are $8$ kernels whose sizes are all set to $2 \times 2$, and the following max pooling size is $2 \times 2$ with strides set to $2$. In the Second convolution layer, there are $16$ kernels whose sizes are all set to $2 \times 2$, and the following max pooling size is the same as the previous max pooling layer. Thus, we get a $1 \times 16$ vector feature for each region proposal. In the sequential aggregation layer, the dimension of LSTM is set to $10$. In the final relevance decision layer, the multi-layer perceptron is a $2$-layer feed forward neural network with one hidden layer whose dimension size is set to $10$. The regularization parameters $\lambda_1$ and $\lambda_2$ are set to $0.0005$ and $0.0001$, respectively. All the other trainable parameters are initialized randomly by uniform distribution within $[-0.1,0.1]$.

\subsubsection{Evaluation Methodology}
Given the limited number of queries for each collection, we conduct 5-fold cross-validation to minimize over-fitting without reducing the number of learning instances. Queries for each dataset are divided into $5$ folds as described in LETOR4.0~\cite{qin2010letor}. The parameters for each model are tuned on 4-of-5 folds. The last fold in each case is used for evaluation. This process is repeated $5$ times once for each fold. The results reported were the average over the $5$ folds.
As for evaluation measures, precision (P), mean average precision (MAP), and normalized discounted cumulative gain (NDCG) at position 1, 5, and 10 were used in our experiments.
We performed significant tests using the paired t-test. Differences are considered statistically significant when the $p-$value is lower than $0.05$.

\begin{table*}[!ht]\centering
\caption{Comparison of different retrieval models over the MQ2007 and MQ2008 datasets. Significant improvement or degradation with respect to our model(ViP with query dependent snapshot) is indicated (+/-) ($\text{p-value} \le 0.05 $).}
\begin{tabular}{l l l l l l l l}
\multicolumn{8}{c}{MQ2007}\\
%\toprule
\hline\hline
Model Name & P@1 & P@5 & P@10 & NDCG@1 & NDCG@5 & NDCG@10 & MAP\\
\hline
BM25 & $0.427^-$ & $0.388^-$& $0.366^-$& $0.358^-$& $0.384^-$& $0.414^-$& $0.450^-$ \\
QL   & $0.401^-$& $0.372^-$& $0.359^-$& $0.347^-$& $0.366^-$& $0.398^-$& $0.430^-$ \\
 \hline
 %\midrule
 RankSVM & $0.472^-$ & $0.413^-$& $0.381^-$& $0.408^-$ & $0.414^-$& $0.442^-$& $0.464^-$\\
 RankBoost & $0.462^-$ & $0.405^-$& $0.374^-$& $0.401^-$ & $0.410^-$& $0.436^-$& $0.457^-$\\
 AdaRank & $0.461^-$ & $0.408^-$& $0.373^-$& $0.400^-$ & $0.415^-$& $0.439^-$& $0.460^-$\\
  LambdaMart & $0.481^-$ & $0.418^-$& $0.384^-$& $0.412^-$ & $0.421^-$& $0.446^-$& $0.468^-$\\
\hline
 ViP & 0.505 & 0.439 & 0.398 & 0.434 & 0.441 & 0.464 & 0.481 \\
 \hline\hline
 \multicolumn{8}{c}{MQ2008}\\
\hline\hline
%\toprule
Model Name & P@1 & P@5 & P@10 & NDCG@1 & NDCG@5 & NDCG@10 & MAP\\
\hline
BM25 & $0.408^-$& $0.337^-$& $0.245^-$& $0.344^-$& $0.461^-$& $0.220^-$& $0.465^-$ \\
QL  & $0.380^-$& $0.323^-$& $0.236^-$& $0.315^-$& $0.441^-$& $0.206^-$& $0.453^-$ \\
 \hline
 %\midrule
 RankSVM &$0.421^-$& $0.350^-$ & $0.247^-$ & $0.357^-$& $0.475^-$ & $0.228^-$ & $0.471^-$\\
 RankBoost & $0.441^-$& $0.347^-$ & $0.248^-$ & $0.368^-$& $0.475^-$ & $0.228^-$~ & $0.478^-$ \\
 AdaRank& $0.434^-$& $0.342^-$ & $0.243^-$ & $0.368^-$& $0.468^-$ & $0.221^-$ & $0.476^-$ \\
 LambdaMart& $0.449^-$& $0.346^-$ & $0.249^-$ & $0.376^-$& $0.471^-$ & $0.230^-$ & $0.478^-$\\
 \hline
 ViP &0.466 & 0.356 & 0.252 & 0.396 & 0.494 & 0.235 & 0.494 \\
 \hline\hline
 %\bottomrule
\end{tabular}
\label{tab:main_results}
\end{table*}

\subsection{Analysis of the ViP model}
In this section, we conduct experiments to analyze the ViP model in learning visual features from snapshots. For this purpose, we introduce two variants of the ViP model. In the first variant, we disable the visual feature learning part and only keep the human-crafted features, referred to as ViP$_\mathit{Baseline}$. In the second variant, we replace the CNN+RNN layer with a widely adopted CNN model~\cite{krizhevsky2012imagenet, bai2017experimental} for image recognition, referred to as ViP$_\mathit{CNN}$. We also test all these models on both query-independent and query-dependent snapshots. The results are shown in Table~\ref{tab:structural_results}.

From the results we observe that, the ViP$_\mathit{Baseline}$ model, which only leverages the human-crafted features, can already obtain reasonably good retrieval performance. 
Furthermore, when snapshots are included, not matter query-independent or query-dependent, the retrieval performance can be significantly improved on both datasets. It indicates that learning visual features from Web page snapshot is of great importance in relevance ranking. Meanwhile, we find that the improvement of query-dependent snapshot is consistently larger than that of the query-independent snapshot in terms of all the evaluation metrics. This is not surprising since the query-dependent snapshots provide richer visual information, i.e., the matching signals between a Web page and a query, than query-independent snapshots. We also try to combine both query-independent and query-dependent snapshots together to learn the visual features, but no obvious improvement can be observed. It indicates that the information in query-independent snapshots might have already been included in query-dependent snapshots.

Moreover, when comparing the ViP model with the ViP$_\mathit{CNN}$ model, we can see that ViP can outperform ViP$_\mathit{CNN}$ consistently in terms of all the evaluation metrics on both query-independent and query-dependent snapshots. For example, the relative improvement of ViP over ViP$_\mathit{CNN}$ on query-dependent snapshot is about  $3.9\%$ and $3.1\%$ in terms of P@1 and NDCG@1 on MQ2007, respectively. The results indicate that the proposed ViP model, which is inspired by the F-biased viewing pattern, can learn the visual features from snapshots more effectively than the existing CNN model which is proposed for general image recognition.

\subsection{Comparison of Retrieval Models}
In this section, we compare our model (ViP model based on query-dependent snapshots) against existing retrieval models over the two benchmark datasets.
The main results are shown in Table~\ref{tab:main_results}.

From the results, we have the following observations: (1) For the two traditional models, we can see that BM25 is a strong baseline which performs better than QL. (2) All the learning to rank models perform significantly better than the traditional retrieval models. It is not surprising since learning to rank models combine various features including the two baseline traditional retrieval models. Among all the learning to rank models, LambdaMart performs best.
(3) We observe that our proposed ViP model can outperform all the existing models in terms of all the evaluation measures on both datasets, and all the improvements are statistically significant (p$-$value$\le 0.05$). For example, On MQ2008 dataset, the relative improvement of our ViP model against the best-performing baseline (i.e. LambdMart) is $3.8\%$, $5.3\%$, and $3.3\%$ with respect to P@1, NDCG@1, and MAP, respectively. The improvement of our model over traditional learning to rank model demontrates the effectiveness of the learned visual features from Web snapshot.
\begin{figure}[tb]
\centering	
\begin{tikzpicture}
	\begin{axis}[
		ybar,
		axis on top,
        height=.32\textwidth,
        width=.5\textwidth,
        bar width=0.44cm,
        ytick={36,38,40,42,44,46,48,50},
        ymajorgrids, tick align=inside,
        major grid style={draw=white},
        minor y tick num={1},
        %tickwidth=0pt,
        enlarge y limits={value=.1,upper},
        ymin=36, ymax=50,
        %axis x line*=bottom,
        %axis y line*=left,
        axis lines=left,
        %y axis line style={opacity=0},
        %tickwidth=0pt,
        enlarge x limits=0.25,
        legend style={
            at={(0.5,1.1)},
            font=\small,
            anchor=north,
            draw=none,
            legend columns=-1,
            /tikz/every even column/.append style={column sep=0.1cm}
        },
        ylabel={\%},
        ylabel style={
            anchor=south,
            at={(ticklabel* cs:1.06)},
            yshift=-3pt
        },
        symbolic x coords={
           P@10,NDCG@10,MAP},
       xtick=data,
       nodes near coords={
        \pgfmathprintnumber[fixed zerofill,precision=1]{\pgfplotspointmeta}
       },
       every node near coord/.append style={anchor=south, font=\fontsize{6pt}{4pt}\selectfont},%\fontsize{4pt}{4pt}\selectfont
	]
	\addplot [draw=none, fill=c2, postaction={pattern=north east lines}] coordinates {
      (P@10,38.9)
      (NDCG@10, 45.6)
      (MAP,47.4)
      };
   \addplot [draw=none,fill=c3, postaction={pattern=vertical lines}] coordinates {
      (P@10,39.8)
      (NDCG@10, 46.4)
      (MAP,48.1)
      };
   \addplot [draw=none, fill=c4, postaction={pattern= crosshatch}] coordinates {
      (P@10,39.2)
      (NDCG@10, 46.0)
      (MAP,47.8)
      };
      \addplot [draw=none, fill=c5, postaction={pattern= crosshatch dots}] coordinates {
      (P@10,38.2)
      (NDCG@10, 45.0)
      (MAP,47.1)
      };
      \legend{P$_{\mathit{2x64}}$,P$_{\mathit{4x64}}$,P$_{\mathit{8x64}}$,P$_{\mathit{16x64}}$}
	\end{axis}

\end{tikzpicture}
\caption{Performance comparison of the ViP model over different proposal sizes on MQ2007.}
\label{fig:proposal_results}
\end{figure}
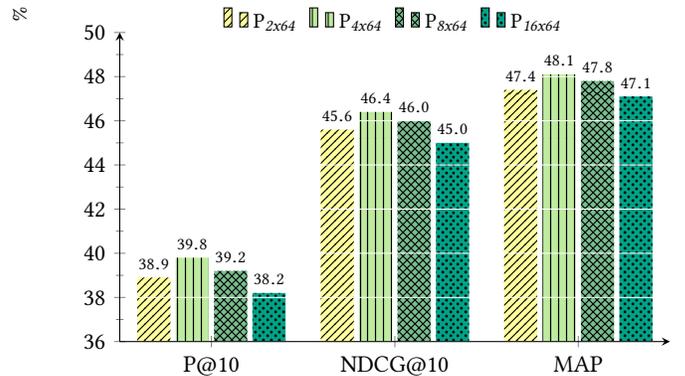

\begin{figure*}[!tbp]
\centering
\includegraphics[scale=0.65]{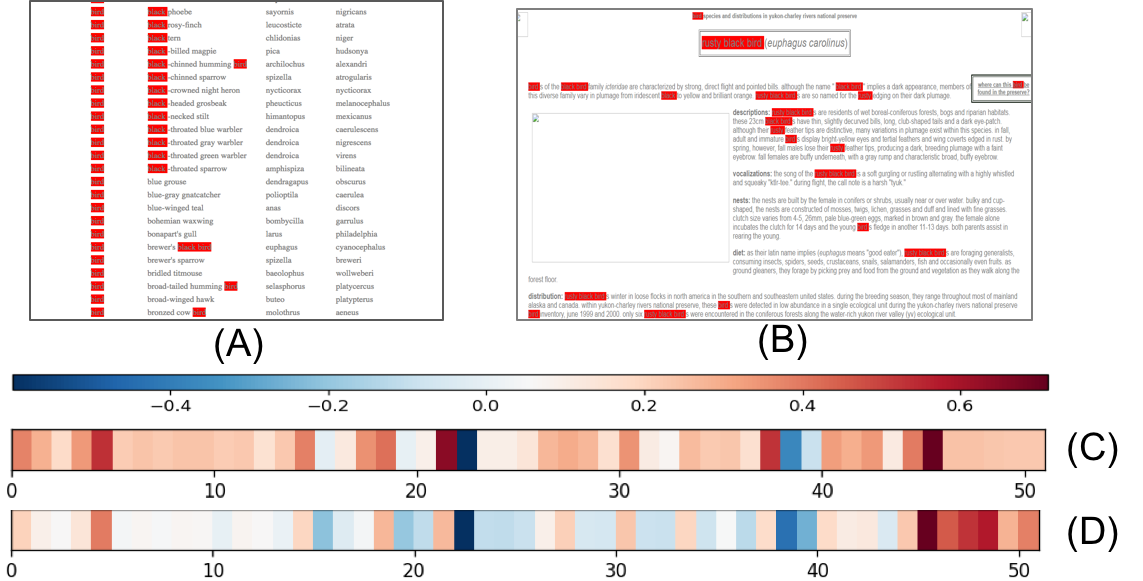}
\caption{(A) The query-dependent snapshot of a non-relevant page for query ``rusty black bird''; (B) The query-dependent snapshot of a highly relevant page for query ``rusty black bird''. (C) The feature importance vector learned by the ViP$_\mathit{Baseline}$ model. (D) The feature importance vector learned by the ViP model.}
\label{fig:case}
\end{figure*}

\subsection{Impact of Proposal Size}
Since we utilize the fixed-height horizontal region proposals of Web snapshot in learning, we would like to study the effect of different heights on the ranking performance. In fact, the proposal height determines not only the granularity of the local information perceived, but also the number of proposals to be aggregated. With a small proposal height, the model can obtain fine-granularity local information, but at the cost of aggregating longer sequence of local features which may require large memories. With a large proposal height, there would be less number of local features to be processed, but it may lose valuable detailed local information.
We conduct experiments to compare different proposal sizes, varying in the range of $2 \times 64$, $4 \times 64$, $8\times 64$, and $16\times 64$. From the results shown in Figure~\ref{fig:proposal_results}, we can see that the performance first increases and then decreases, with the increase of proposal size. The best performance is obtained when the proposal size is set to be $4\times 64$ (w.r.t. different evaluation measures). This height approximately corresponding to two lines of text in the original Web page in the normal font size.

\subsection{Impact of Image Resolution}
As is described in Section 4.1, we down-sample a Web snapshot to a lower spatial resolution with $64 \times 64$ pixels for efficiency. As we know, different snapshot resolutions preserve different amount of information, which may affect the visual features learned by the ViP model.
Here we analyze the effect of different snapshot resolutions, varying in the range of $16 \times 16$, $32 \times 32$, $64 \times 64$, $128 \times 128$, and $256 \times 256$ pixels. Note that in the previous section, we find that the performance could also be affected by the size of region proposal. Thus, for snapshots with different resolution, we also tune the best proposal size. We find the best proposal size under different resolutions is $2\times 16$, $2\times 32$, $4 \times 64$, $8\times 128$, and $16 \times 256$, respectively. 
Therefore, we compare the performance over different snapshot resolutions under the best-performing region proposal size, and the results are depicted in Figure~\ref{fig:resolution}. Moreover, we also plotted the performance results of the ViP$_\mathit{Baseline}$ model for additional comparison.

From the results we can see that, the performance increases rapidly with the snapshot resolution, and then keeps stable after a certain point. When the resolution is low, the performance of the ViP model may even decrease as compared with the ViP$_\mathit{Baseline}$ model which only uses human-crafted features. The possible reason might be that the snapshot with too small resolution may lose useful information of the Web page, leading to undesired noise in learning. We observe that the medium resolution with size $64 \times 64$ can already obtain significant performance improvement. This is quite important since we only need to store relatively small snapshots for effective usage in real search application.

\pgfplotsset{
axis background/.style={fill=white},
grid=both,
  xtick pos=left,
  ytick pos=left,
  tick style={
    major grid style={style=gallery,line width=1pt},
    minor grid style=mercury,
    },
  minor tick num=1,
}

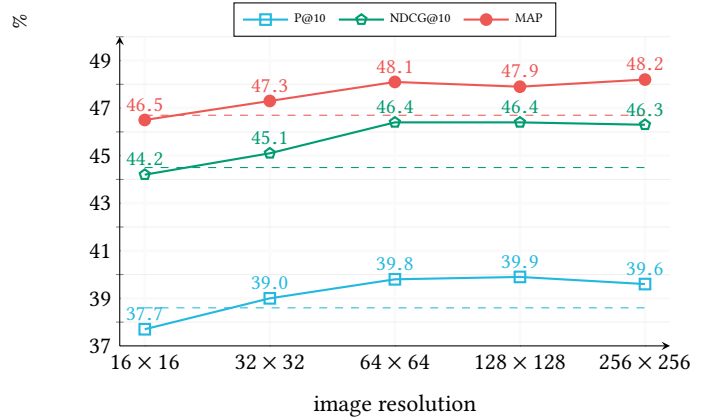
\begin{figure}[!tb]
\centering
  \begin{tikzpicture}
    \begin{axis}[
      height=.32\textwidth,
      width=.5\textwidth,
      %xlabel={passage size},
      %every axis y label/.style={ at={(ticklabel* cs:1.05)}, anchor=south },
      xticklabels={$16\times 16$,$32 \times 32$,$64 \times 64$,$128 \times 128$,$256 \times 256$},
      xtick={1,2,3,4,5},
      legend style={
          font=\tiny,
          legend columns=-1,
          at={(0.5,1)},
          anchor=south,
          /tikz/every even column/.append style={column sep=0.9mm}
        },
        ymajorgrids={true},
        ymin=37,
        ymax=50,
        ytick={37, 39, ..., 49},
        %major grid style={draw=white},
        minor x tick num={0},
        minor y tick num={1},
        %y axis line style={opacity=80},
        %axis y line* = left,
        %axis x line* = bottom,
        axis lines=left,
        enlarge x limits=0.05,
        tickwidth=0pt,
        %nodes near coords,
        every node near coord/.append style={anchor=north, font=\fontsize{8pt}{4pt}\selectfont},%\fontsize{4pt}{4pt}\selectfont
        legend entries = {P@10, NDCG@10, MAP},
        xlabel={image resolution},
        ylabel={\%},
      	ylabel style={
      	      anchor=south,
            at={(ticklabel* cs:1.05)},
            yshift=-3pt
        },
        %every tick label/.append style={font=\scriptsize},
        %nodes near coords={
        %\pgfmathprintnumber[fixed zerofill,precision=1]{\pgfplotspointmeta}
       	%}
        ]
	 %\tikzset{elegant/.style={smooth,samples=5, magenta}}
	  %\addplot[elegant, domain=xtick]{45.3};

      \addplot[color=shakespeare,mark=square,
          nodes near coords={ \pgfmathprintnumber[fixed zerofill,precision=1]{\pgfplotspointmeta}},
          nodes near coords style={anchor=south},
          thick] coordinates {
        (1, 37.7)
        (2, 39.0)
        (3, 39.8)
        (4, 39.9)
        (5, 39.6)
      };
      \addplot[color=free_speech_aquamarine, mark=pentagon,
          nodes near coords={ \pgfmathprintnumber[fixed zerofill,precision=1]{\pgfplotspointmeta}},
      	  nodes near coords style={anchor=south},
          thick] coordinates {
        (1, 44.2)
        (2, 45.1)
        (3, 46.4)
        (4, 46.4)
        (5, 46.3)
      };
      \addplot[color=flamingo, mark=*,
      	 nodes near coords={ \pgfmathprintnumber[fixed zerofill,precision=1]{\pgfplotspointmeta}},
      	 nodes near coords style={anchor=south},
      	 thick] coordinates {
        (1, 46.5)
        (2, 47.3)
        (3, 48.1)
        (4, 47.9)
        (5, 48.2)
      };
      \addplot[dashed, color=shakespeare] coordinates {
        (1, 38.6)
        (2, 38.6)
        (3, 38.6)
        (4, 38.6)
        (5, 38.6)
      }; 
      \addplot[dashed, color=free_speech_aquamarine] coordinates {
        (1, 44.5)
        (2, 44.5)
        (3, 44.5)
        (4, 44.5)
        (5, 44.5)
      };
      \addplot[dashed, color=flamingo] coordinates {
        (1, 46.7)
        (2, 46.7)
        (3, 46.7)
        (4, 46.7)
        (5, 46.7)
      };
    \end{axis}
  \end{tikzpicture}
  \caption{Performance comparison over snapshots under different resolutions on MQ2007.}
  \label{fig:resolution}
\end{figure}

\subsection{Case Study}

To better understand what can be learned by the ViP model, here we conduct some case studies. Figure~\ref{fig:case} shows two candidate Web pages of the query ``rusty black bird'' in MQ2008 dataset, which have totally different layouts. The page (A) (DocId: GX059-61-15727287 in GOV2 corpus) shows a list of vertebrate animal species list, and is labeled as Non-Relevant to the query. The page (B) (DocId: GX095-93-12495293 in GOV2 corpus) describes the rusty black bird in detail, and is labeled as Highly Relevant to the query. When we apply the ViP$_\mathit{Baseline}$ model over the query, we find page (A) at the second position in the ranking list. The possible reason is that there are more than 300 hits of the query keyword "bird" in this page, and the ViP$_\mathit{Baseline}$ model thus may produce a high relevance score by mainly relying on human-crafted textual features. On the other hand, the page (B), which has fewer query term matches, is ranked at a lower position (i.e., the seventh position). However, if we take into account the layout of the Web page, we can clearly see that the matching signals in page (A) are distributed vertically along with many blank areas, indicating a table or list of elements in a Web page. While in page (B), there are many matching signals embedded in passages closely with some large matching singles in the top position (i.e., title), indicating a descriptive article in the Web page. By taking these visual features into account, the ViP can better detect the relevant Web page, and promote page (B) to the second position and penalize page (A) to the nineteenth position in the ranking list.

Moreover, we also depict the learned weights in the feedforward layer to analyze the feature importance. For better visualization and analysis, we simplify the decision layer in our model by using only one-layer feedforward neural network. In this way, the weights form a vector with the size of the total features. As shown in Figure~\ref{fig:case}, the Figure~\ref{fig:case}(C) is the learned weights of the ViP$_\mathit{Baseline}$ model, and Figure~\ref{fig:case}(D) is the learned weight of the ViP model. In both Figures, the color is corresponding to the signal strength, where red represents the  highest value. From Figure~\ref{fig:case}(C) we find that the most important features are BM25 score of the body and the number of child page. However, when the visual features are included, the weights change significantly. From Figure~\ref{fig:case}(D) we observe that many visual features become very important, while the importance of many human-crafted features, especially those link analysis features,  decreases significantly. For example, the weight of PageRank, inlink number, and outlink number decreased $5.5\%$, $39\%$, and $54.5\%$, respectively (i.e. from 41 to 43). This is reasonable since many link analysis features also convey page quality information, which can now be well captured by visual features from Web snapshots.

\section{Conclusions}
In this paper, we propose to learn visual features from Web page snapshots to improve the performance of ad-hoc Web retrieval. Both query-independent and query-dependent snapshots have been introduced as new inputs. We then propose a novel visual perception model over the snapshots, which can automatically learn visual features in an end-to-end way.  Experimental results on two benchmark datasets have demonstrated that visual features from Web snapshots can significantly improve the performance of ad-hoc Web retrieval. We have also shown that this method can be efficiently implemented in practical search systems with an efficient indexing scheme. For future work, it would be interesting to apply our visual perception model to other Web page related applications, e.g., spamming detection, homepage identification or mobile Web search.

\section{Acknowledgments}
This work was funded by the 973 Program of China under Grant No. 2014CB340401, the National Natural Science Foundation of China (NSFC) under Grants No. 61232010, 61433014, 61425016, 61472401, and 61203298, the Youth Innovation Promotion Association CAS under Grants No. 20144310 and 2016102, and the National Key R\&D Program of China under Grants No. 2016QY02D0405. We would like to thank Zhicheng Dou for the Web page rendering tool. 

\bibliographystyle{ACM-Reference-Format}

\end{document}